\documentclass[aps,pre,twocolumn,floatfix,showpacs]{revtex4-1}
\usepackage{graphicx}
\usepackage{dcolumn}
\usepackage{bm}
\usepackage{color}
\usepackage{natbib}
\usepackage{array}
\usepackage{subfigure}
\usepackage{amssymb}
\usepackage{amsbsy}
\begin{document}
    \title{Lagrangian velocity auto-correlations in statistically-steady rotating turbulence}
    \author{Lorenzo Del Castello}
    \author{Herman J.H. Clercx}
    \affiliation{Department of Physics and J.M. Burgers Centre for Fluid Dynamics, Eindhoven University of Technology, P.O. Box 513, 5600 MB Eindhoven, The Netherlands}
    \date{\today}
    \begin{abstract}
		Lagrangian statistics of passive tracers in rotating turbulence is investigated by Particle Tracking Velocimetry. A confined and steadily-forced turbulent flow is subjected to five different rotation rates. The PDFs of the velocity components clearly reveal the anisotropy induced by background rotation. Although the statistical properties of the horizontal turbulent flow field are approximately isotropic, in agreement with previously reported results by van Bokhoven and coworkers [Phys. Fluids 21, 096601 (2009)], the velocity component parallel to the (vertical) rotation axis gets strongly reduced (compared to the horizontal ones) while the rotation is increased. The auto-correlation coefficients of all three components are progressively enhanced for increasing rotation rates, although the vertical one shows a tendency to decrease for slow rotation rates.
		The decorrelation is approximately exponential. Lagrangian data compare favourably with previously reported Eulerian data for horizontal velocity components, but show different behaviour for the vertical velocity components at higher rotation rates. 
    \end{abstract}
    %
    %\pacs{47.27.-i,47.32.Ef}
    %
    \maketitle
    \section{Introduction}
	The influence of the rotation of the Earth on oceanic and atmospheric currents, as well as the effects of a rapid rotation on the flow inside industrial machineries like mixers, turbines, and compressors, are only the most typical examples of fluid flows affected by rotation. Despite the fact that the Coriolis acceleration term appears in the Navier-Stokes equations with a straightforward transformation of coordinates from the inertial system to the rotating non-inertial one, the physical mechanisms of the Coriolis acceleration are subtle and not fully understood yet. Several fluid flows affected by rotation have been studied by means of direct numerical simulations (DNS) and analytical models.
	For example, DNS studies addressing the role of rotation on velocity correlations and mixing~\cite{yeung1998pof,yeung2004}, the role of vertical confinement~\cite{godeferd1999}, energy spectra in (decaying) rotating turbulence~\cite{smith1999,smith2005,chen2005,mininni2009}, and scaling laws in rotating turbulence~\cite{mueller2007} have been reported. Several experimental studies of rotating turbulence have been carried out~\cite{ibbetson1975,hopfinger1982,jacquin1990,baroud2002,baroud2003,morize2005,morize2006,davidson2006,staplehurst2008,bokhoven2009,moisy2010}. However, quantitative experimental data is rather scarce and purely of Eulerian nature~\cite{bokhoven2009,moisy2010}.\\
	The present work addresses experimentally the topic, focusing on a class of fluid flows of utmost importance: confined and continuously forced rotating turbulence. In recent experimental investigations on (decaying) rotating turbulence quantitative information is extracted by means of Particle Image Velocimetry (PIV)~\cite{moisy2010} and stereo-PIV~\cite{bokhoven2009}; the present investigation is based on Particle Tracking Velocimetry (PTV), thus acquiring Lagrangian statistics of rotating turbulence for the first time in laboratory settings.\\
	A useful insight into the structure of a turbulent flow field is represented by the auto-correlations of the velocity field in the Lagrangian frame. The integral time scales derived from the Lagrangian velocity correlations give a rough estimate of the time a fluid particle remains trapped inside a large-scale eddy, and therefore it might be used as a lower-bound for the typical lifetime of the large eddies. Lagrangian correlations of velocity have been recognised as the key-ingredient of the process of turbulent diffusion since the work by Taylor \cite{taylor1921plmsa,monin1975sfm}. Since then, the Lagrangian view-point received a growing attention, see for a recent review Ref.~\cite{toschi2009}.
	Lagrangian correlations of velocity in non-rotating turbulence were recently measured with an acoustic technique at very high Taylor-based Reynolds number ($Re_{\lambda}\simeq800$) and a decay of the correlation coefficients of single velocity components proportional to $e^{-\tau/\tau_0}$ was proposed, with $\tau_0$ comparable to the energy injection time scale~\cite{mordant2001prl,mordant2004njp}. The same decay has been observed by Gervais {\it{et al.}}~\cite{gervais2007eif}, who compared Eulerian and Lagrangian correlations of velocity in a $Re_{\lambda}\simeq320$ turbulent flow, also relying on acoustic measurements. Here, some of these issues are addressed for rotating turbulence as measured by means of PTV.
    \section{Experimental setup}
	\begin{figure}
	    \includegraphics[width=0.485\textwidth]{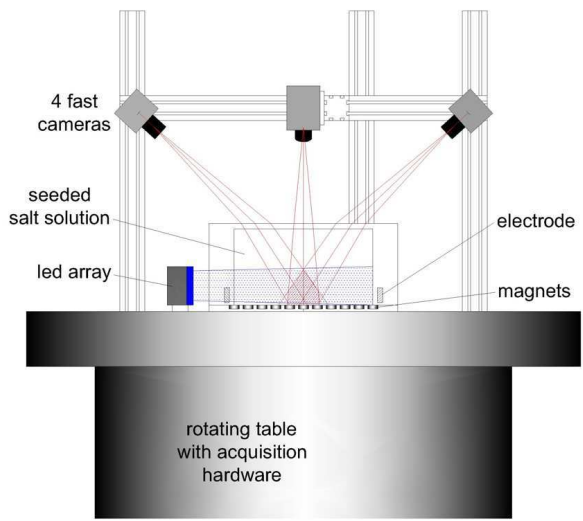}
	    \caption{(Colour online) Schematic drawing of the experimental setup, side view. A perspex container sits on top of a rotating table, and is filled with a NaCl solution. An array of permanent magnets is placed below the container, and two linear electrodes are immersed in the fluid. An aluminium frame holds four digital cameras in stable position (three of them are visible in the drawing), and their common field-of-view is sketched. A powerful LED-array, on the left of the container, provides the necessary illumination.}
	    \label{fig1}
	\end{figure}
	\begin{figure}
	    \includegraphics[width=0.30\textwidth]{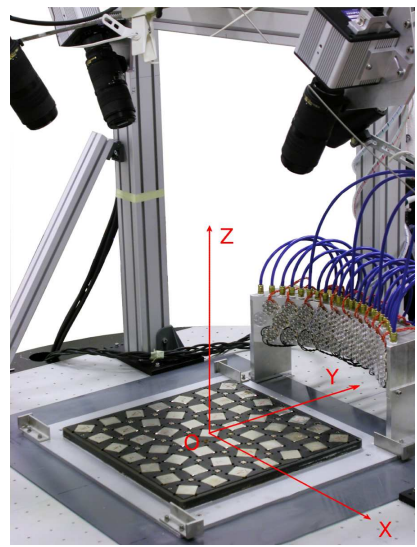}%{setup02e_xyz_s.eps}
	    \caption{(Colour online) Picture of the setup partially mounted: the full magnet array is surrounded by the light source (on the right) and the cameras (on top). The full array of $7\times7$ large magnets, and the smaller magnets placed between them, are visible. The position of the Cartesian reference frame $\{x,y,z\}$ is indicated by the (red) arrows.}
	    \label{fig2}
	\end{figure}
	The experimental setup consists of a fluid container, made of transparent perspex in order to ensure optical accessibility, equipped with a turbulence generator, and an optical measurement system. A side-view of the setup is sketched in Fig.~\ref{fig1}, and a photograph of the setup partially mounted is shown in Fig.~\ref{fig2}. Four digital cameras (Photron FastcamX-1024PCI, three of them partially visible in Fig.~\ref{fig2}) acquire images of the central-bottom region of the flow domain through the top-lid. The fluid is illuminted by means of a LED-array made of $238$ Luxeon K2 LEDs ($1.4~\mathrm{kW}$ total dissipation and roughly $150~\mathrm{W}$ of light) mounted on a thick aluminium block provided with water-cooling channels. The illumination system and its cooling connections are visible in Fig.~\ref{fig2}. These key elements are mounted on a rotating table, so that the flow is measured in the rotating frame of reference.
	The inner dimensions of the container define a flow domain of $500\times500\times250~\mathrm{mm^3}$ (length $\times$ width $\times$ height); note that the free surface deformation is inhibited by a perfectly sealed top lid. The turbulence generator is an adaptation of a well-known electromagnetic forcing system commonly used for shallow-flow experiments~\cite{sommeria1986,tabeling1991,dolzhanskii1992}, and currently operational in our laboratory for both shallow flow and rotating turbulence experiments~\cite{clercx2003,akkermans2008,bokhoven2009}. The tank is filled with a highly concentrated sodium chloride (NaCl) solution in water, 28.1\% brix (corresponding to 25 g NaCl in 100 g of water). The fluid density $\rho_{f}$ is $1.19~\mathrm{g/cm^3}$ and the kinematic viscosity $\nu$ is $1.319~\mathrm{mm^2/s}$. Two titanium elongated electrodes are placed near the bottom at opposite sidewalls of the container.
	A remote-controlled power supply (KEPCO BOP 50 8P) is connected to the electrodes and provides a stable electric current of $8.39~\mathrm{A}$.
	An array of axially magnetized permanent (neodymium) magnets is placed directly underneath the bulk fluid. Fig.~\ref{fig2} reveals the array of magnets, before the fluid container is mounted on the table. The magnets have a magnetic field strength of approximately 1.4 T at the centre of the magnet surface, and they are arranged following a chessboard scheme, {\it{i.e.}} alternating North and South poles for the magnet’s top faces. The magnets, kept in position by a polyvinyl chloride (PVC) frame, are fixed on a 10 mm thick steel plate to increase the density of the magnetic field lines in the fluid bulk. A range of flow scales is forced by using two differently sized magnets, viz., i) elongated bar magnets, $10\times10\times20~\mathrm{mm^3}$ in size; and ii) flat bar magnets, $40\times40\times20~\mathrm{mm^3}$ in size~\cite{bokhoven2009}. With such arrangement, the largest scales that are forced are comparable with the spacing between adjacent large magnets, \textit{i.e.} $\mathcal{L}^F=70~\mathrm{mm}$.\\
	The Lagrangian correlations are measured by means of Particle Tracking Velocimetry, making use of the code developed at ETH, Z\"{u}rich~\cite{maas1993eif1,malik1993eif2,willneff2002istpdrm,willneff2003phd,luthi2005jfm}. PMMA (poly methyl methacrylate) particles, with a mean diameter $d_p=127\pm3\mathrm{\mu m}$ and particle density $\rho_p=1.19~\mathrm{g/cm^3}$, are used as flow tracers. The concentration of the salt solution is adjusted to match the PMMA density. The Stokes number ($St$) for these tracers expresses the ratio between the particle response time and a typical time scale of the flow. For the present experiments it can be estimated as $St=\tau_p/\tau_{\eta}={\mathcal{O}}(10^{-3})$, where $\tau_p = d_p^2/(18\nu)$ is the particle response time (with $\rho_{p}/\rho_f=1$) and $\tau_{\eta}$ is the Kolmogorov time scale of the turbulent flow, which values are given in the following section.
	The chosen seeding particles can thus be considered as passive flow tracers both in terms of buoyancy and inertial effects. An accurate calibration of the measurement system on a 3D-target, followed by the optimisation of the calibration parameters on seeded flow images, permits to retrieve the 3D-positions of the particles with a maximum error of $9~\mathrm{\mu m}$ in the horizontal directions, and $18~\mathrm{\mu m}$ in the vertical one. The data is then processed in the Lagrangian frame, where the trajectories are filtered to remove the measurement noise produced by the positioning inaccuracy: third-order polynomials are fitted along limited segments of the trajectories around each particle position (for details, see Ref.~\cite{luthi2002phd}). From the coefficients of the polynomial in each point, the 3D time-dependent signals of position and velocity are extracted.
	With the present setup, up to 2500 particles per time-step have been tracked on average in a volume with size $100\times100\times100~\mathrm{mm^3}$, thus roughly $1.5\mathcal{L}^F$ along each coordinate direction.\\
	A detailed description of the experimental setup and the data processing routines, together with an in-depth characterisation of the flow, can be found in \cite{delcastello2010phd}.
    \section{Characterisation of the flow}
\begin{figure}
	    \includegraphics[width=0.48\textwidth]{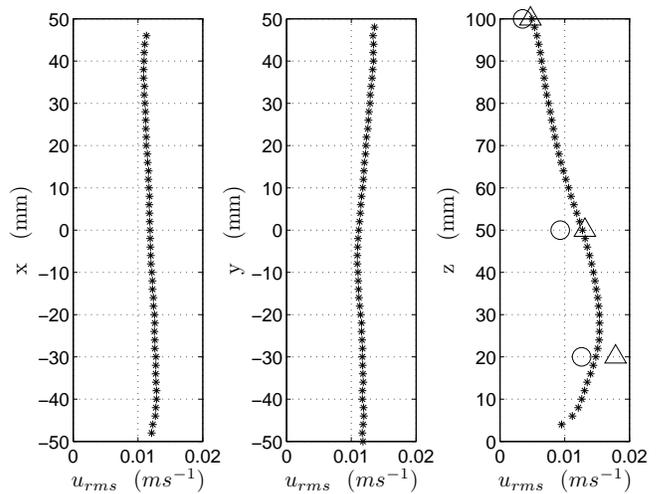}%{081212ptvmed_exp01bis-0rad_Uefluctrms-xyzprof_480dt.eps}
	    \caption{Profiles of the r.m.s. velocity magnitude $u_{rms}\equiv\langle u^2\rangle^{1/2}$ averaged over planes perpendicular to the coordinate direction under consideration, for the non-rotating experiment. The large symbols on the $z$-profile indicate the corresponding values as measured with stereo-PIV during previous experimental campaigns (circles: $4.00~\mathrm{A}$; triangles: $8.00~\mathrm{A}$ forcing current).}
	    \label{fig3}
	\end{figure}

\begin{figure}
	    \includegraphics[width=0.48\textwidth]{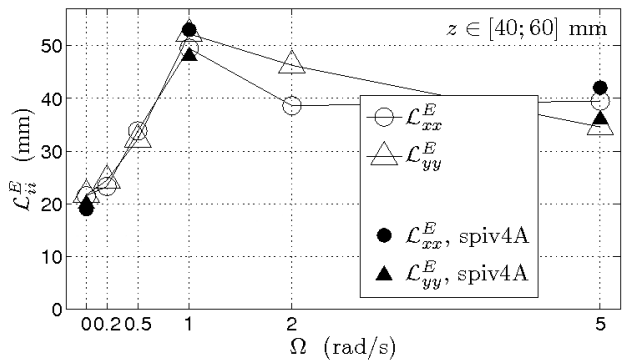}%{exps123456_espacecorrv3-fstep10-zslab3-integralscales_z-removed.eps}
	    \caption{Eulerian longitudinal integral length scales $\mathcal{L}^E_{ii}$ in the horizontal directions against the rotation rate $\Omega$ for the mid-height horizontal slice, $z\in[40;60]~\mathrm{mm}$ (open symbols). The corresponding horizontal lenght scales measured by means of stereo-PIV at $z=50~\mathrm{mm}$ are also indicated on the plot (solid symbols).}
	    \label{fig4}
	\end{figure}

	The flow is subjected to different background rotation rates $\Omega\in\lbrace0; 0.2; 0.5; 1.0; 2.0; 5.0\rbrace~\mathrm{rad/s}$ around the vertical $z$-axis. The measurements are performed when the turbulence is statistically steady (measured by the kinetic energy of the flow). The mean kinetic energy of the turbulent flow is then constant in time and decays in space along the upward vertical direction. The flow is fully turbulent in the bottom region of the container where the measurement domain is situated. Eulerian characterisation of the (rotating) turbulent flow with stereo-PIV measurements has been reported elsewhere~\cite{bokhoven2009}.\\
	The values of important flow quantities in the measurement domain are reported in Table~\ref{tab1}: the root-mean-square (r.m.s.) of each velocity component $u_{i,rms}\equiv\langle u_i^2\rangle^{1/2}$ and the ratio of horizontal and vertical values $\xi\equiv0.5(u_{x,rms}+u_{y,rms})/u_{z,rms}$; the Rossby number $Ro\equiv u_{rms}/(2\Omega\mathcal{L}^F)$; the Ekman number $Ek\equiv\nu/(\Omega L_z^2)$; the thickness of the Ekman boundary layer $\delta_{Ek}\equiv\sqrt{\nu/\Omega}$.
	\linespread{1.3} % it corresponds to linespacing 1.5
	\begin{table}%[!h]
	    \begin{center}
		\begin{tabular}{cccccccc}
		    \hline
		    $\Omega$ \footnotesize{(rad/s)} & & 0 & 0.2 & 0.5 & 1.0 & 2.0 & 5.0\\
		    \hline
		    Root mean square~~~~
			& $x$ & 9.6 & 9.4 &  9.8 & 12.0 & 17.0 & 14.4 \\
		    $\langle u_i^2\rangle^{1/2}$,~~~~~\small{with $i=$}
			& $y$ & 9.6 &  9.1 &  9.8 & 12.1 & 17.5 & 12.2 \\
		    \footnotesize{($\mathrm{mm/s}$)}
			& $z$ & 8.3 &  7.7 &  7.8 &  6.6 &  7.3 &  2.2 \\
		    \hline
		    $\xi~\footnotesize{(-)}$  &    & 0.86 & 0.83 & 0.80 & 0.55 & 0.42 & 0.17 \\
		    $Ro~\footnotesize{(-)}$  &    & $\infty$ & 0.47 & 0.20 & 0.13 & 0.09 & 0.02 \\
%		    $Ek~\footnotesize{(-)}$  &    & $\infty$ & $1\times10^{-4}$ & $4\times10^{-5}$ & $2\times10^{-5}$ & $1\times10^{-5}$ & $4\times10^{-6}$\\
		    $Ek\times10^{5}~\footnotesize{(-)}$  &    & $\infty$ & $10$ & $4$ & $2$ & $1$ & $0.4$\\
		    $\delta_{Ek}~\mathrm{(mm)}$ & & $\infty$ & 2.5 & 1.6 & 1.1 & 0.8 & 0.5\\
		    \hline
		\end{tabular}
		\linespread{1}
		\caption{Root mean square (r.m.s.) values of the components of velocity; ratio of horizontal and vertical r.m.s. values $\xi\equiv0.5(u_{x,rms}+u_{y,rms})/u_{z,rms}$; Rossby number $Ro\equiv u_{rms}/(2\Omega\mathcal{L}^F)$; Ekman number $Ek\equiv\nu/(\Omega L_z^2)$, with $L_z=250~\mathrm{mm}$ the vertical size of the flow domain; thickness of the Ekman boundary layer $\delta_{Ek}\equiv\sqrt{\nu/\Omega}$, for each experiment.}
		\label{tab1}
	    \end{center}
	\end{table}
	It is noteworthy to emphasize the higher value of the r.m.s. of the velocity components for $\Omega=2.0~\mathrm{rad/s}$: this anomalous behaviour may be connected with instabilities of large-scale anticyclonic vortical structures (see, e.g., Refs.~\cite{kloosterziel1991jfm,heijst2009arfm}) at this rotation rate, to be expected for Rossby close to the critical value $0.1$ ($Ro\simeq0.2$ in similar experiments by Hopfinger {\it{et al.}}~\cite{hopfinger1982}). Such instabilities are under further investigation. Furthermore, the strong suppression of vertical velocity at the maximum rotation rate $\Omega=5.0~\mathrm{rad/s}$ represents a classical signature of fast rotation, \textit{i.e.} the two-dimensionalisation of the flow field. The transition to 2D, in a first approximation, can be quantified in terms of the ratio $\xi$.
	Despite the anomaly observed for $\Omega=2.0~\mathrm{rad/s}$, the ratio $\xi$ is monotonically decreasing with increasing rotation rate $\Omega$, indicating that the two-dimensionalisation process proceeds despite the probable occurrence of anticyclonic instabilities. The Ekman number varies from $10^{-4}$ to $4\times10^{-6}$, and the Ekman viscous boundary layer has negligible thickness. For the Kolmogorov length and time scales we found the typical values $0.6~{\rm{mm}}\lesssim\eta\lesssim0.8~{\rm{mm}}$ and $0.25~{\rm{s}}\lesssim\tau_{\eta}\lesssim0.55~{\rm{s}}$, respectively.
	The Taylor-scale Reynolds number is in the range $70\lesssim Re_{\lambda}\lesssim110$ for all rotation rates, except for $\Omega = 2.0~\mathrm{rad/s}$, for which a larger value is found.\\
	In order to investigate the horizontal homogeneity of the forced flow field in case of no rotation, and to quantify the vertical inhomogeneity, profiles of the r.m.s. of the velocity magnitude are plotted in the three directions, and shown in Fig.~\ref{fig3}. The flow appears to be homogeneous to a good approximation in the horizontal directions.
	On the vertical profile, the corresponding values obtained via stereo-PIV measurements (van Bokhoven {\it{et al.}}~\cite{bokhoven2009}) on three horizontal planes are also reported for comparison. Circles represent the values obtained from experiments with a lower forcing intensity ($4.00~\mathrm{A}$ in place of $8.39~\mathrm{A}$ used for PTV experiments); triangles come from experiment with almost the same forcing settings ($8.00~\mathrm{A}$), and a very good agreement is observed for these runs between PTV and stereo-PIV measurements. Such agreement is also supported by an almost perfect match between Eulerian horizontal longitudinal integral length scales from stereo-PIV and PTV measurements for the range of rotation rates considered~\cite{delcastello2010phd}, which are shown in Fig.~\ref{fig4}.
	Rotation induces a significant increase of the horizontal lenght scales up to $1~\mathrm{rad/s}$, and a decrease for faster rotations, in excellent agreement with the stereo-PIV measurements. The data by van Bokhoven {\it{et al.}}~\cite{bokhoven2009} also show that the flow is approximately isotropic at mid-height in the measurement domain, an important result which can be and is used in the analysis of the present data.
    \section{Results}
	We present and discuss here the PDFs of velocity and the Lagrangian auto-correlations of velocity as obtained from the described experiment.
	\subsection{Probability distribution functions of velocity}
	    We first report on the PDFs of velocity (each one computed on roughly $4\times 10^6$ data points).
	    \begin{figure}
			\includegraphics[width=0.35\textwidth]{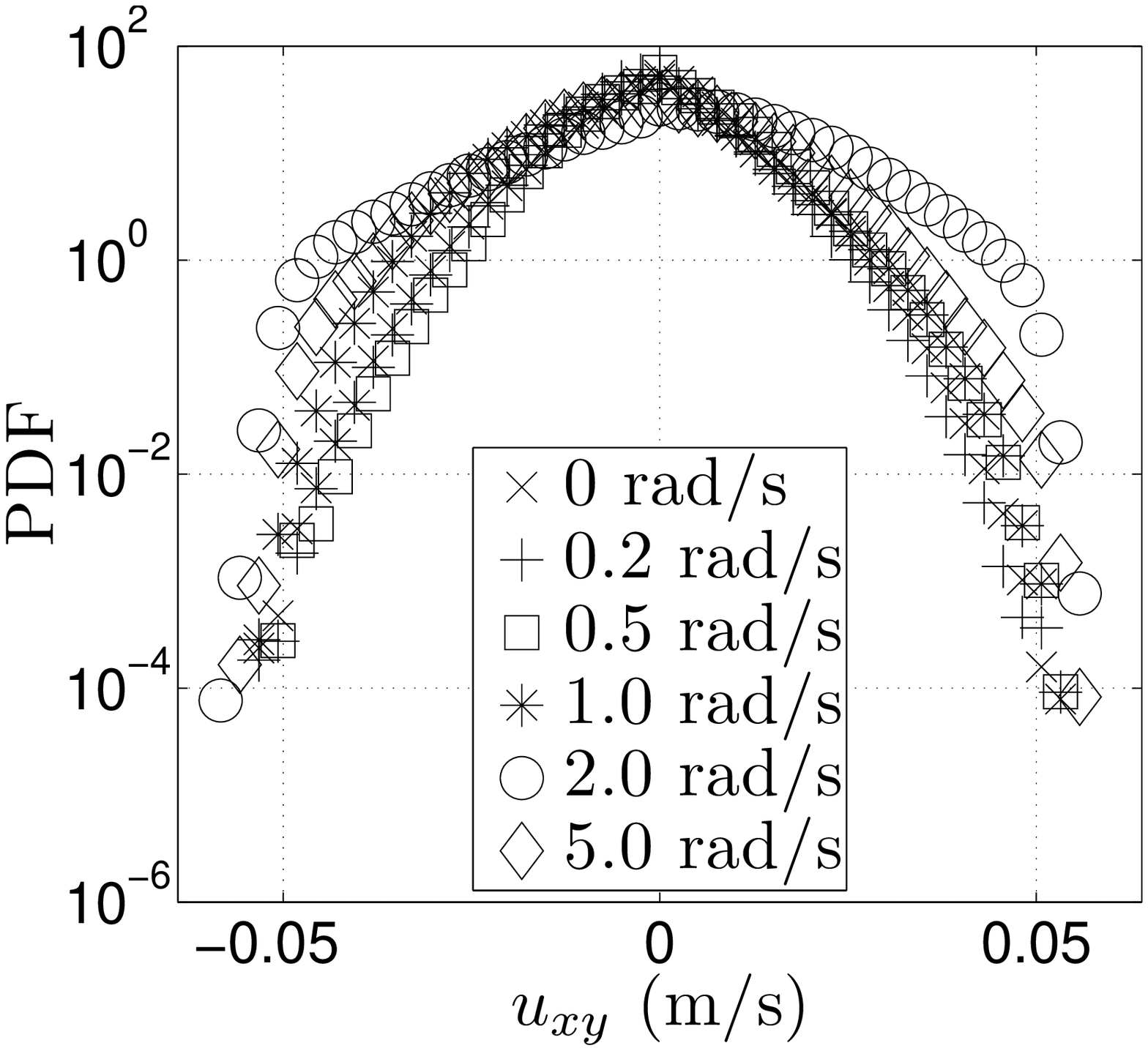}\\%{exps123456_PDFuxy-fstep5-linlog.eps}\\
			\includegraphics[width=0.35\textwidth]{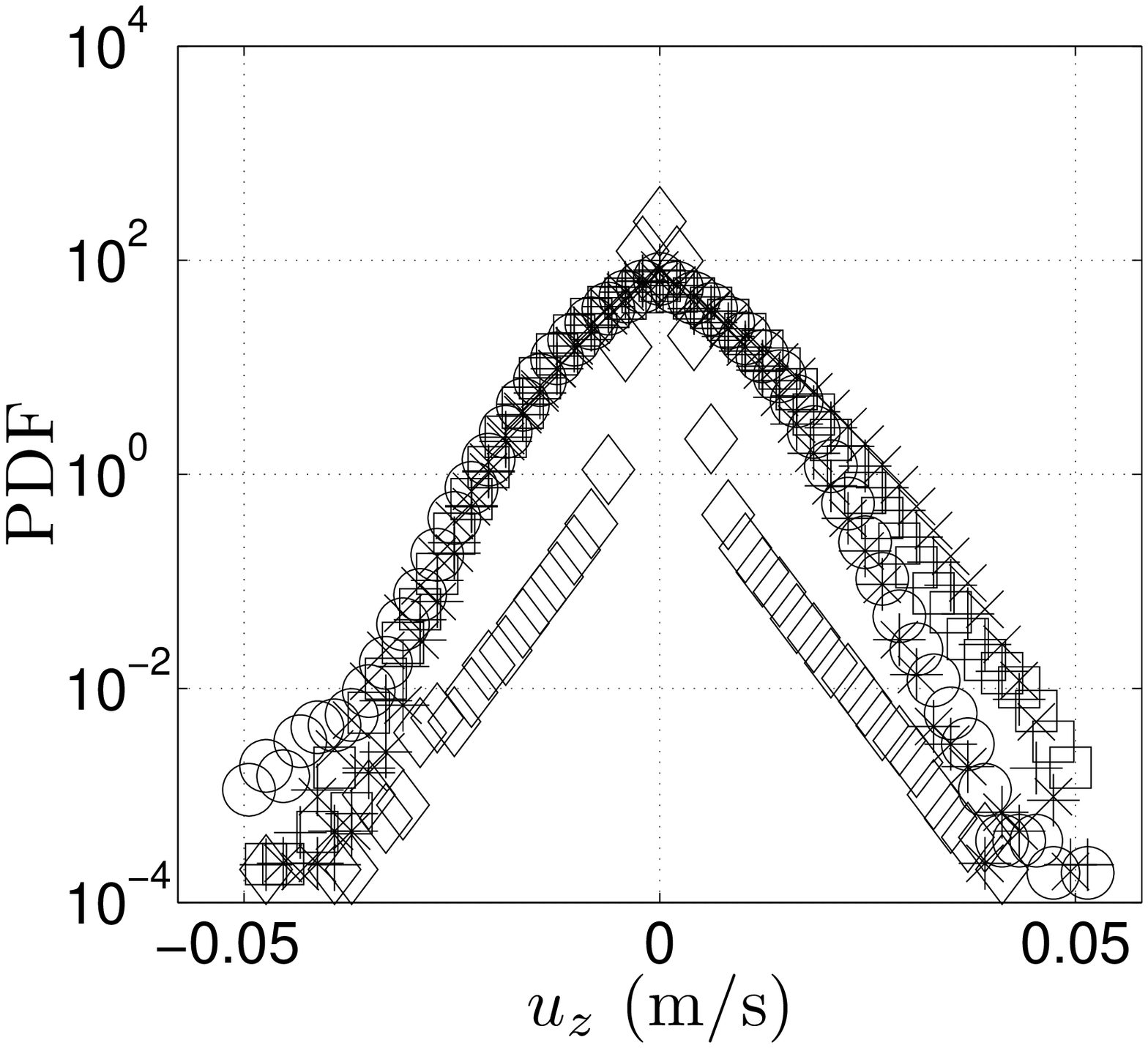}%{fig2c.eps}
			\caption{PDFs of the velocity components for all experiments, in linear-logarithmic scale. Top-panel: average of the PDFs of the two horizontal components. Bottom-panel: PDF of the vertical component. The two plots share the same legend.}
			\label{fig5}
		\end{figure}
	    The PDFs are shown in Fig.~\ref{fig5} in linear-logarithmic scale for all experiments together. Using the assumptions of horizontal homogeneity and isotropy, the PDFs of the $x$- and $y$-component are averaged together and shown in the top-panel of the figure; in the bottom-panel, the PDF of the $z$-component is reported. The background rotation is seen to induce only a slight anisotropy of the horizontal components of velocity (the PDFs for $\Omega=2.0~\mathrm{rad/s}$ clearly reflect a larger $u_{rms}$). The most important effect of rotation is seen on the vertical velocity component, for which the standard deviation of the PDF gets strongly damped for $\Omega=5.0~\mathrm{rad/s}$. The distributions for $\Omega\in\lbrace0;1.0;5.0\rbrace~\mathrm{rad/s}$ are in good quantitative agreement with the ones published by van Bokhoven {\textit{et al.}} \cite{bokhoven2009} (see Figs. $8$ and $14$ therein).
	    The PDFs have in both cases almost Gaussian shapes (minor skewness, except for $\Omega=2.0~\mathrm{rad/s}$) and the kurtosis is only slightly larger than the Gaussian value. We found $3.0\lesssim\langle u_i^4\rangle/\langle u_i^2\rangle^2\lesssim4.0$, except for the vertical velocity component at $\Omega=5.0~\mathrm{rad/s}$ which shows a substantially larger value for the kurtosis. Once more, the latter describes the well-known effect of rotation, which suppresses the fluid motion in the direction of the rotation axis, hence inducing a strong 2D-character of the flow field.
	\subsection{Lagrangian velocity auto-correlations}
\begin{figure}
			\includegraphics[width=0.42\textwidth]{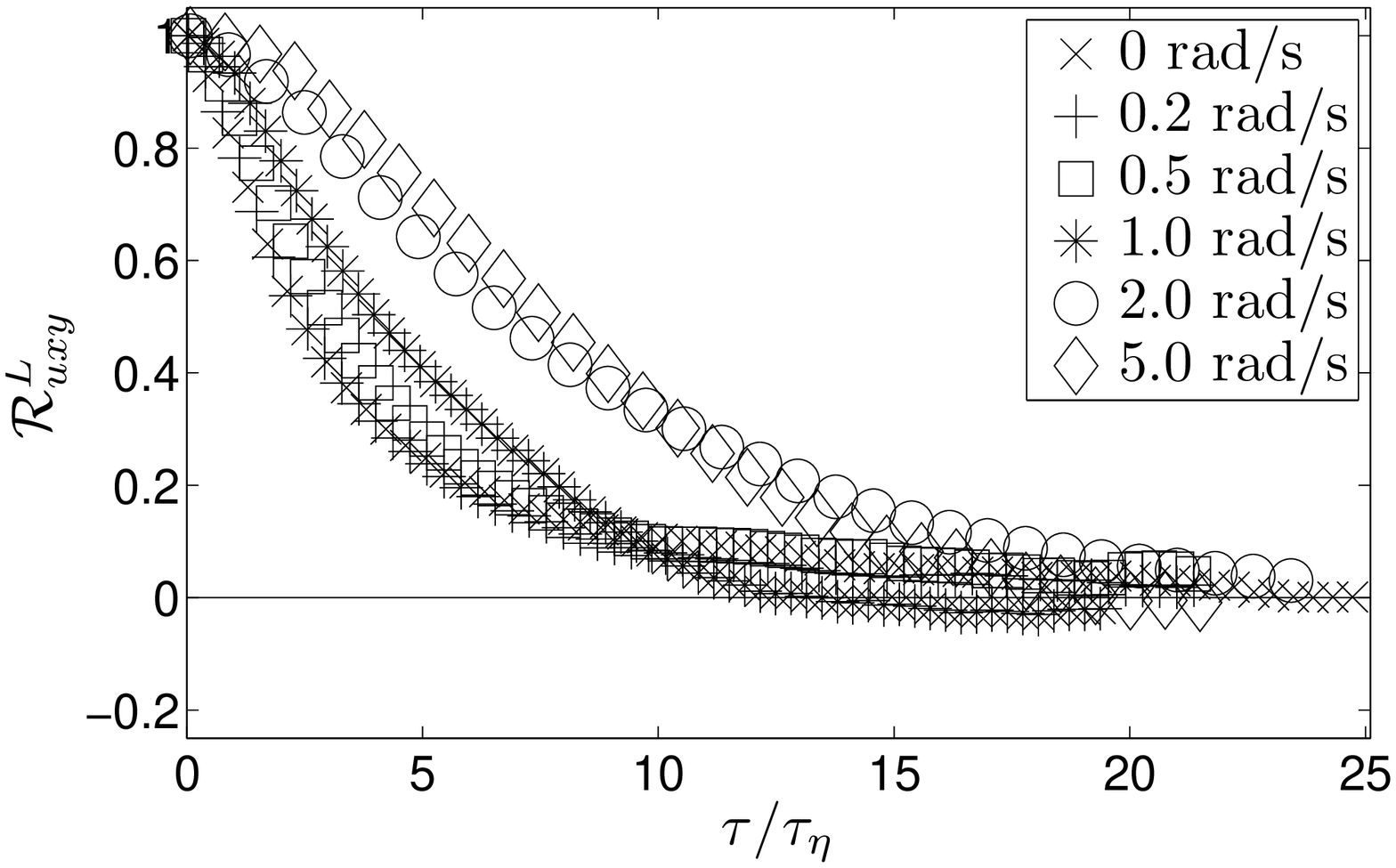}\\%{exps123456_lagrvelacccorr-v11-minlife50-uxy-intaueta-linlin.eps}\\
			\includegraphics[width=0.42\textwidth]{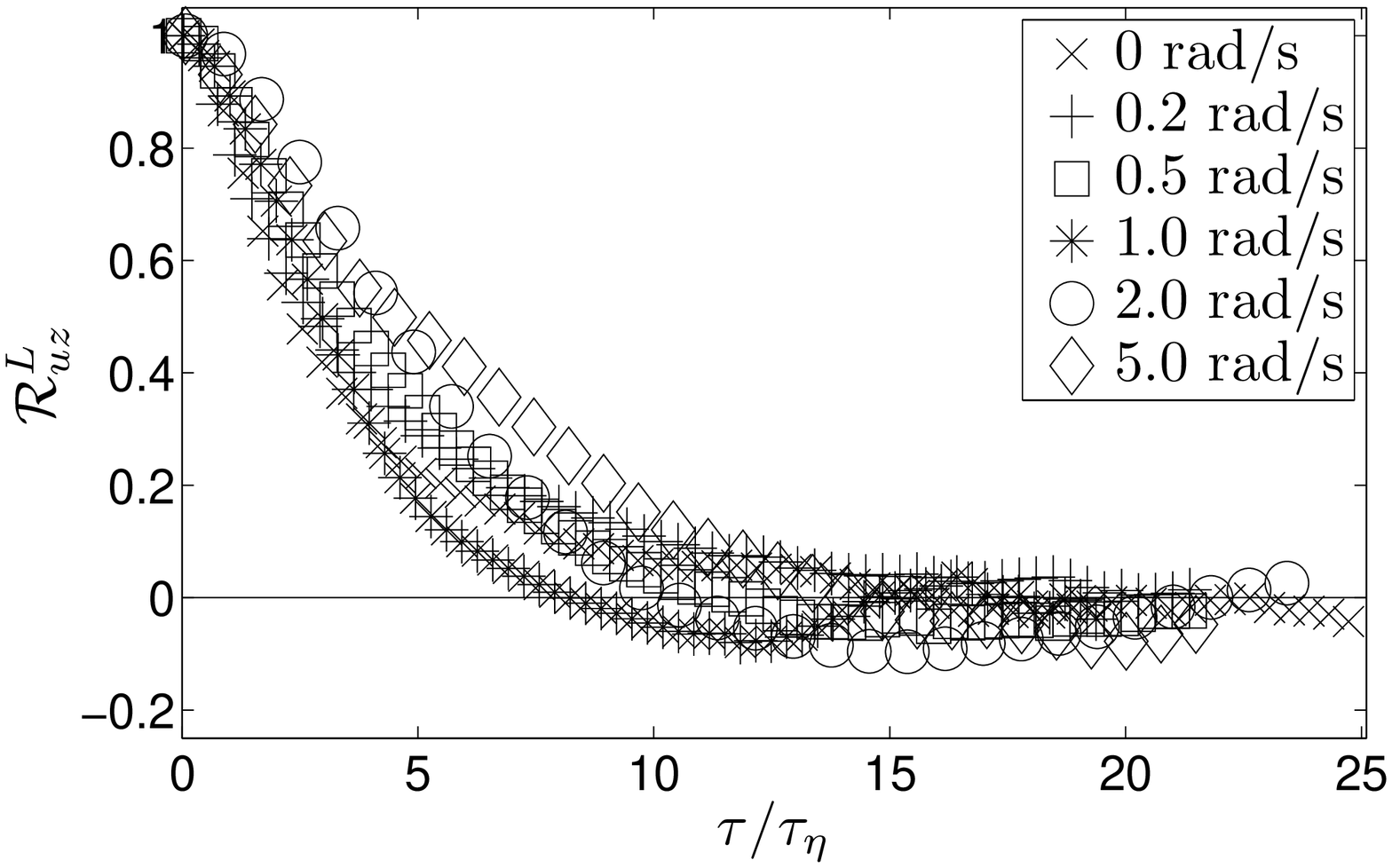}%{fig3c.eps}
			\caption{Lagrangian auto-correlation coefficients of velocity for all rotating experiments, with time normalised with the Kolmogorov time scale $\tau_{\eta}$. Top-panel: average of the correlation coefficients of the horizontal components $u_x$ and $u_y$. Bottom-panel: correlation coefficient of the vertical component $u_z$. Only one symbol every ten data points is plotted for readability.}
			\label{fig6}
	    \end{figure}

\begin{figure}
			\includegraphics[width=0.48\textwidth]{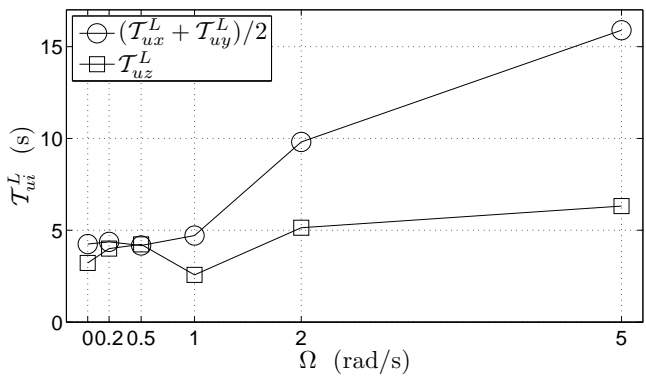}
			\caption{Horizontal and vertical Lagrangian integral time scales against the rotation rate $\Omega$. The values are estimated as the constant $\tau_0$ of the exponential fit $e^{-\tau/\tau_0}$ performed over the correlation curves shown in  Fig.~\ref{fig6}, for the time interval over which each curve shows a convincing exponential decay. The two horizontal time scales are averaged together and compared with the vertical time scale.}
			\label{fig7}
	    \end{figure}

	    The auto-correlation coefficients $\mathcal{R}^L_{ui}(\tau)$ for each velocity component $u_i(t)$ (with $i\in{1,2,3}$ denoting the $x$-, $y$- and $z$-component, respectively), which are functions of the time separation $\tau$, are obtained by averaging over a sufficient number of trajectories, and normalising with the variance of the single component, i.e.:
	    \begin{equation}
			\mathcal{R}^L_{ui}(\tau)\equiv\frac{\langle u_i(t)u_i(t+\tau)\rangle}{\langle u_i^2(t)\rangle}~.
	    \end{equation}
	    It is also useful to define the three associated integral time scales:
	    \begin{equation}
			\mathcal{T}^L_{ui}\equiv\int_{0}^{\infty}\mathcal{R}^L_{ui}(\tau)\mathrm{d}\tau~.
	    \end{equation}
	    For strongly anisotropic turbulence, as the one influenced by fast background rotation, the individual scales in directions parallel and perpendicular to the rotation axis may differ substantially. Their comparison permits to quantify the anisotropy of the large-scale flow.\\
	    The Lagrangian auto-correlation coefficients of velocity for all experiments are shown in Fig.~\ref{fig6} with time $\tau$ normalised with the Kolmogorov time $\tau_{\eta}$. The top-panel shows the average of the correlation coefficients of the two horizontal components of velocity; the bottom-one shows the correlation coefficient of the vertical velocity component.
	    For times separations $\tau$ longer than $10\tau_{\eta}$, some of the correlations show a non-perfect statistical convergence, which is due to the limited recording time available with the present camera system (equipped with onboard RAM memory). Despite this, the correlations describe clearly a monotonic influence of rotation: the coefficients gets progressively higher for increasing $\Omega$, both for the horizontal components and for the vertical one. Additionally, a stronger Lagrangian auto-correlation is found for the vertical velocity component (relative to those of the horizontal components) than previously reported for the Eulerian temporal velocity correlations~\cite{bokhoven2009}. The linear-logarithmic plots reveal that the decorrelation is roughly exponential, at least till the coefficients drop under $0.4$, in good agreement with the relevant literature (see, e.g., Refs.~\cite{mordant2001prl,mordant2004njp,gervais2007eif}).
	    The exponential decay of the velocity auto-correlation plays an essential role in some dispersion models, strongly characterising them~\cite{sawford1991pofa}. Following \cite{mordant2001prl}, we fit the function $e^{-\tau/\tau_0}$ over all curves, limited to the time interval over which each curve shows a convincing exponential decay. The Lagrangian integral time scales $\mathcal{T}^L_{ui}$ are then estimated as the constant $\tau_0$ retrieved from each fit. In order to facilitate the comparison between horizontal and vertical time scales, we average together the two horizontal scales, in view of the symmetry of our flow around the $x$- and $y$-axis (horizontal isotropy). The results are plotted against the rotation rate and compared with the vertical time scale in Fig.~\ref{fig7}.
	    Despite the limited accuracy of the estimated Lagrangian integral time scales, the trends summarised in Fig.~\ref{fig7} reflect the results shown in Fig.~\ref{fig6} and allow an easier quantification of the process. The horizontal scale progressively increases with increasing rotation rate (and are of similar size as the earlier reported Eulerian integral time scales~\cite{bokhoven2009}). The vertical one, on the contrary, shows a tendency to decrease slightly for $\Omega$ up to $1.0~\mathrm{rad/s}$, and increases only for higher rotation rates.
	    \\
    \section{Conclusions}
	We set up an experiment to investigate the statistical properties of a continuously-forced statistically-steady turbulent flow subjected to different background rotation rates. The flow was first characterised in terms of the velocity r.m.s., Rossby and Ekman numbers. The profiles in the three coordinate directions were inspected, revealing the horizontal homogeneity of the flow, and describing the vertical decay of energy due to the increasing distance from the forcing system. The data obtained with a different measurement system in similar experiments (with access to the full range of vertical velocity fluctuations in a horizontal plane)~\cite{bokhoven2009} confirm the same vertical decay of energy, and reveal that the flow is isotropic at mid-height in the measurement domain.
	The good agreement between the two datasets was also shown in terms of the Eulerian horizontal integral length scales, which are seen to increase substantially for mild background rotation rates and to decrease slightly for higher rotation rates. The influence on the PDFs of the velocity components is shown, revealing the two-dimensionalisation process induced by rotation. We then used the PTV data to explore the auto-correlations of the velocity components in the Lagrangian frame, in order to quantify the memory of the velocity of fluid parcels along their trajectories. All auto-correlation coefficients are progressively enhanced for increasing rotation rates, although the vertical one first decreases slightly for slow rotation rates.
	The decorrelation process is found to be approximately exponential. Comparison of the Lagrangian data with the Eulerian measurements from similar rotating turbulence experiments~\cite{bokhoven2009} suggests that fluid parcels, being restricted to coherent flow structures, have limited access to vertical velocity variations when the rotation rate is increased. Eulerian measurements would over-estimate the sampling over vertical velocity fluctuations. This is particularly shown by the enhanced memory in the Lagrangian vertical velocity auto-correlation compared to its Eulerian counterpart.\\\\
    \section*{Acknowledgements}
	This project has been funded by the Netherlands Organisation for Scientific Research (NWO) under the Innovational Research Incentives Scheme grant ESF.6239. The Institute of Geodesy and Photogrammetry and the Institute of Environmental Engineering of ETH (Z\"{u}rich) are acknowledged for making available the PTV code. The authors would like to thank Beat L{\"u}thi and Arkady Tsinober for the useful scientific discussions.
\end{document}